\documentclass[reqno,a4paper]{amsart}

\usepackage{caption}
\usepackage{booktabs} 
\usepackage{amsmath, latexsym, amsthm, amsfonts, bm, amssymb} 
\usepackage{graphicx}
\usepackage{url}
\usepackage{float}
\usepackage{enumerate}
\usepackage[foot]{amsaddr}
\usepackage{textcomp} 
\usepackage{xcolor}
\usepackage{tikz}
\PassOptionsToPackage{hyphens}{url}\usepackage{hyperref}
\usepackage{psfrag,epsfig,epstopdf}

\usepackage{natbib}
\bibpunct{(}{)}{;}{a}{}{,} 

\allowdisplaybreaks

\theoremstyle{plain}

\newtheorem{exam}{Example}

\theoremstyle{remark}

\renewcommand{\hat}{\widehat}

\newcommand{\x}{{\rm x}}

\newcommand{\rR}{\mathbb R}

\def\bSig\mathbf{\Sigma}

\renewcommand{\hat}{\widehat}

\newcommand{\rd}{{\rm d}}

\definecolor{lime}{HTML}{A6CE39}
\DeclareRobustCommand{\orcidicon}{%
	\begin{tikzpicture}
	\draw[lime, fill=lime] (0,0) 
	circle [radius=0.16] 
	node[white] {{\fontfamily{qag}\selectfont \tiny ID}};
	\draw[white, fill=white] (-0.0625,0.095) 
	circle [radius=0.007];
	\end{tikzpicture}
	\hspace{-2mm}
}

\foreach \x in {A, ..., Z}{%
	\expandafter\xdef\csname orcid\x\endcsname{\noexpand\href{https://orcid.org/\csname orcidauthor\x\endcsname}{\noexpand\orcidicon}}
}

\begin{document}

\title[{S}eparable nonlinear least-squares]{Separable nonlinear least-squares parameter estimation for complex dynamic systems}

\author[I. Dattner]{Itai Dattner$^1$$^\mathsection$\orcidA{}}
\address{$^1$Department of Statistics,
	University of Haifa,
	199 Aba Khoushy Ave.,
	Mount Carmel,
	Haifa 3498838,
	Israel}
\email{idattner@stat.haifa.ac.il}
\thanks{$^\mathsection$To whom correspondence should be addressed.}

\author[S.~Gugushvili]{Shota Gugushvili$^2$\orcidB{}}
\address{$^2$Biometris,
	Wageningen University \& Research,
	Postbus 16,
	6700 AA Wageningen,
	The Netherlands}
\email{shota.gugushvili@wur.nl}
\thanks{The research leading to the results in this paper has received funding from the European Research Council under ERC Grant Agreement 320637. Itai Dattner was supported by the Israeli Science Foundation grant number 387/15.}

\author[H. Ship]{Harold Ship$^1$}
\email{harold.ship@gmail.com}

\author[E. O. Voit]{Eberhard O. Voit$^3$\orcidD{}}
\address{$^3$The Wallace H. Coulter Department of Biomedical Engineering at Georgia Institute of Technology and Emory University, 950 Atlantic Drive, Atlanta, GA, 30332--2000, USA}
\email{eberhard.voit@bme.gatech.edu}



\begin{abstract}
Nonlinear dynamic models are widely used for characterizing functional forms of processes that govern complex  biological pathway systems. Over the past decade, validation and further development of these models became possible due to data collected via high-throughput experiments using methods from molecular biology. While these data are very beneficial, they are typically incomplete and noisy, so that inferring parameter values for complex dynamic models is associated with serious computational challenges. Fortunately, many biological systems have embedded linear mathematical features, which may be exploited, thereby improving fits and leading to better convergence of optimization algorithms. 

In this paper, we explore options of inference for dynamic models using a novel method of {\it separable nonlinear least-squares optimization}, and compare its performance to the traditional nonlinear least-squares method. The numerical results from extensive simulations suggest that the proposed approach is at least as accurate as the traditional nonlinear least-squares, but usually superior, while also enjoying a substantial reduction in computational time.
\end{abstract}

\maketitle

\section{Introduction}
Nonlinear dynamic models are widely used for characterizing functional forms of  processes that govern complex biological pathway systems. Of particular interest in this context are so-called canonical formats, which are very flexible in their possible responses, yet involve a very restricted domain of functional forms. Outside linear systems, the best-known canonical formats are Lotka-Volterra (LV) models (\cite{lotka1956}, \cite{volterra1926}, \cite{peschel1986predator}, and \cite{may2001}), which use binomial terms, and power-law systems within the framework of Biochemical Systems Theory (BST), which exclusively use products of power functions. BST was originally devised for the analysis of biochemical and gene regulatory systems, but has subsequently found much wider application in various biomedical and other areas (\cite{savageau1976} and \cite{voit2013biochemical}). Whereas it is easy to set up an LV or BST model for a complex biological system in a symbolic format, the identification of optimal parameter values continues to be a significant challenge. As a consequence, estimating parameters of systems of ordinary differential equations (ODEs) remains to be an active research area that attracts contributions from a variety of scientific fields (e.g., \cite{gennemark2007}, \cite{chou2009recent}, \cite{zbMATH07045664}, \cite{mmp2012dsreview}, \cite{ramsay17}, and \cite{schittkowski02}). Indeed, numerous optimization methods for ODE models have been proposed in recent years, but none works exceptionally well throughout wide ranges of application. 

The main focus of this paper is {\it not} a new estimation method {\it per se}. Instead, we are interested in a more general and high-level point of view regarding parameter estimation.
Specifically, our work addresses parameter estimation for dynamic models whose mathematical form contains significant linear features, that allow a natural separation of parameters and system states. A trivial example is a linear ODE where the vector field $x^\prime(t)=\theta x(t)$ is linear in the parameter $\theta$, with $x^\prime(t)$ denoting the derivative of $x(t)$ with respect to $t$. As a more interesting example, the ODE vector field may be partially linear in the parameters, as it is the case for so-called S-system models in BST.
\begin{exam}
An S-system (see \cite{voit2000computational}) is defined as
\begin{equation}\label{eq:s-system}
x^\prime_j(t)=\alpha_j\prod_{k=1}^dx_k^{g_{jk}}(t)-\beta_j\prod_{k=1}^dx_k^{h_{jk}}(t), \quad j=1\dots,d.
\end{equation}
Here $\alpha_j,\beta_j$ are positive rate constants, while $g_{jk},h_{jk}$ are real-valued kinetic orders that reflect the strength and directionality of the effect that a variable has on a given influx or efflux. Informally, one can view this system as a regression equation, where the ``covariates'' are the variables $x_j(t)$ on the righthand side, whereas the ``response'' variables are the derivatives $x_j^{\prime}(t)$ on the left-hand side. Note that the vector field is linear in the rate constants  $\alpha_j,\beta_j$, but nonlinear in the kinetic orders $g_{jk},h_{jk}$. \qed
\end{exam}

Estimation methods that exploit separability of parameters and system states in dynamic models have a long history; see, e.g., \cite{himmelblau1967determination} for a special case. However, a rigorous statistical analysis of one such method has been achieved  only recently (\cite{dattner2015}). In a classical paper on the inference for dynamic models, \cite{varah1982spline} mentions in passing that ``one can use the idea of separability or variable projection (see \cite{golub1973differentiation}  or \cite{ruhe1980algorithms}), in which the linear parameters are implicitly solved for, the resulting (fully) nonlinear least squares problem is solved for the nonlinear parameters, and then the linear parameters are obtained using their representation in terms of the nonlinear parameters. Since this reduces the size of the nonlinear least squares problem to be solved, it is worthwhile.'' Somewhat surprisingly, given that parameter estimation for ODEs is commonly thought as a computational bottleneck in modeling dynamic processes, Varah's suggestion has \emph{not} been widely followed in practice. In fact, in the vast literature dedicated to parameter fitting techniques for dynamic models, we are aware of only two relevant references: \cite{dattner2017modelling}, using the direct integral approach, applied the separable nonlinear least-squares to the inference of parameters in a predator–prey system acting in a heterogeneous environment, while \cite{wu2018parameter} used separability to estimate parameters of high-dimensional linear ODE systems. Moreover, Varah's idea of exploiting separability for estimation of ODE parameters has been implemented only recently in a publicly available software package (\cite{simode}). 

The analysis in this paper is hoped to convince the reader that Varah's idea is indeed worth pursuing. Specifically, we explore and compare two general data fitting approaches for dynamic models: the traditional nonlinear least-squares method (NLS), and a proposed separable nonlinear least-squares method (SLS). Through extensive Monte-Carlo simulations of representative complex models, arising in different scientific fields, we identify and quantify significant statistical and computational gains when using this separation method. We will ultimately come to the conclusion that model separability can be very beneficial and that the SLS approach should be considered for complex dynamic systems with significant linear features.

The paper is organized as follows. In Section \ref{sec:sep} we present details of the SLS methodology in the context of dynamic models. Section \ref{sec:exp} describes the simulation setup, quantifies the statistical measures we use in order to compare the performance of SLS and NLS, and presents numerical results, while conclusions are provided in Section \ref{sec:con}.

\section{Separable nonlinear least-squares (SLS) and Varah's idea}\label{sec:sep}

\subsection{Generalities}
Following Varah's original idea within the context of inference in dynamic models, the main advantages of exploiting separability for parameter estimation are the following (\cite{golub2003separable}):
\begin{enumerate}[(i)]
	\item fewer initial guesses are required for optimization,
	\item the optimization problem is better conditioned, and
	\item convergence is faster. 
\end{enumerate}
These advantages have been convincingly demonstrated in several publications. For example, see \cite{mullen2008separable} for an implementation and applications in physics and chemistry; \cite{chung2010efficient} for a high-dimensional case, where the number of parameters is substantially larger than the number of observations; \cite{gan2017some}, who compared the performance of several algorithms for SLS problems; and \cite{erichson2018sparse}, who studied sparse principal component analysis via variable projection. Separable models are of broad practical applicability, and as such form a subject of active theoretical and applied research. For instance, when analyzing the ``reduced'' nonlinear optimization problem of a separable structure, simplified conditions are required for establishing a variety of theoretical results concerning numerical and statistical properties of the resulting estimators, compared to the original NLS problem (e.g., \cite{basu2000stability} and \cite{dattner2015}). 

In the following we focus on complex dynamic models and, specifically, consider a system of ordinary differential equations given by
\begin{equation}\label{eq:ode_model}
\bigg\{
\begin{array}{l}
 x^{\prime}(t)= F( x(t);\theta),\ t\in[0,T],
\\
 x(0)=\xi,
\end{array}
\end{equation}
where $ x(t)$ takes values in $\rR^d,$ $\xi\in\Xi\subset \rR^d,$
and $\theta\in\Theta\subset\rR^p.$ Let
\begin{equation}\label{eq:sep_ode}
 F( x(t);\theta)= g(x(t);\theta_{NL})\theta_L,
\end{equation}
where $\theta=(\theta_{NL}^\top,\theta_{L}^\top)^\top$, and the symbol $\top$ stands for the matrix transpose. Here $\theta_{NL}$, a vector of size $p_{NL}$, stands for the ``nonlinear'' parameters that are not separable from the state variables $x$, while $\theta_{L}$, a vector of size $p_L$, contains the ``linear'' parameters; note that $p=p_L+p_{NL}$. As the vector field in \eqref{eq:sep_ode} is separable in the linear parameter vector $\theta_L$, we refer to the corresponding ODE system as {\it linear in the parameter}  $\theta_L$ (cf.\ the case of a linear regression model), although the {\it solution} to the system might be highly nonlinear in $\theta$, or even implicit.

\begin{exam}
Let \[\theta_{NL}=(g_{11},\dots,g_{1d},\dots,g_{d1},\dots,g_{dd},h_{11},\dots,h_{1d},\dots,h_{d1},\dots,h_{dd})^\top,\]
and $\theta_L=(\alpha_1,\beta_1,\dots,\alpha_d,\beta_d)^\top.$ Then one sees that equation \eqref{eq:s-system} is a special case of \eqref{eq:ode_model}--\eqref{eq:sep_ode}. \qed
\end{exam}

\subsection{Solution strategy}
Let $x(t; \theta,\xi),$  $t \in [0, T ],$ be the solution of the initial value problem \eqref{eq:ode_model}. We assume that noisy measurements $Y_{j}(t_i)$ on the system are collected at time points $t_i \in [0,T]$, where
\begin{equation}
\label{eq:obs}
Y_{j}(t_i)=x_j(t_i; \theta,\xi)+\epsilon_{ij}, \quad i=1,\ldots,  n, \quad j=1,\ldots,d.
\end{equation}
Here the random variables $\epsilon_{ij}$ are unobservable, independent
measurement errors (not necessarily Gaussian) with zero mean and finite variance.

Varah's approach to parameter estimation in ODE models works as follows. Let $\hat{x}(t)$  stand for a smoother of the data, obtained, e.g., using splines or local polynomials (see, e.g., \cite{fan1996}, \cite{green1994}, and \cite{wasserman06nonparametric} for a treatment of various smoothing methods and an extensive bibliography). This smoother approximates the solution $x(t;\theta,\xi)$ to the ODE \eqref{eq:ode_model}. Varah suggests to insert the smoother into equation \eqref{eq:ode_model}, which will now be satisfied only approximately, and to minimize the resulting discrepancy over the parameters $\xi$ and $\theta.$ A minimizer $(\hat{\xi},\hat{\theta})$ is then an estimator of $(\xi,\theta).$ This idea was put on a solid statistical foundation in \cite{brunel2008parameter} and \cite{gugushvili2012sqrt}. Varah's original approach requires the use of the derivative $\hat{x}^{\prime}(t)$ as an estimator of $x^{\prime}(t),$ which is a disadvantage, as it is well-known that estimating derivatives from noisy and sparse data may be rather inaccurate: see, e.g., \cite{vilela2007} and \cite{chou2009recent}, or more generally \cite{fan1996}. Recent research (\cite{dattner2015},  \cite{dattner2015model}, \cite{vujavcic2015time}, \cite{chen2017network}, \cite{mikkelsen2017learning}, \cite{dattner2017modelling}, \cite{yaarietal18}, \cite{dattnergugushvili18}, \cite{yaari2018textbf}, \cite{dattner2018modern}) has shown that it is more fruitful to transplant Varah's idea to the solution level of equation \eqref{eq:ode_model}. The corresponding approach is referred to as integral estimation and proceeds as follows. 
Define the integral criterion function
\begin{equation}
\label{eq:int_crit}
\int_0^T\left|\left|
\hat{x}(t)-\xi-\int_0^t F(\hat x(s);\theta)\,\rd s\right|\right|^2\rd t,
\end{equation}
where $\left|\left|\cdot\right|\right|$ is the Euclidean norm.
A minimizer of \eqref{eq:int_crit} over $(\xi,\theta)$ gives a parameter estimator. In practice, the integral has to be discretized and replaced by a sum, so that minimization can be performed using the nonlinear least-squares method, 
\begin{equation}\label{eq:nls}
(\hat\xi_{NLS},\hat\theta_{NLS})=\arg\min_{\xi,\theta}\int_0^T\left|\left|
\hat{x}(t)-\xi-\int_0^t F(\hat x(s);\theta)\,\rd s\right|\right|^2\rd t.
\end{equation}
The NLS solution does not take into account the specific linear form of the ODEs in \eqref{eq:sep_ode}, but uses the general form in \eqref{eq:ode_model}.

It is at this stage that Varah suggested to utilize separability, without actually investigating such an approach. However, details are easy to work out. Denote
\begin{eqnarray*}\label{eq:Ghat}
	\hat{G}(t) &:=&\hat{G}(t;\theta_{NL})= \int_0^t g(\hat{x}(s);\theta_{NL})\,\rd s\,,\quad t \in
	[0,T],\nonumber \\
	\hat{A} &=& \int_0^T \hat{G}(t)\,\rd t, \\
	\hat{B} &=& \int_0^T \hat{G}^\top(t) \hat{G}(t)\,\rd t. \nonumber
\end{eqnarray*}
Then, with $\theta_{NL}$ kept fixed, a minimizer of \eqref{eq:int_crit} is given by
\begin{eqnarray*}
	\hat{\xi}(\theta_{NL}) &=& \left(TI_d - \hat{A} \hat{B}^{-1}
	\hat{A}^\top\right)^{-1} \int_0^T \left(I_d - \hat{A}
	\hat{B}^{-1}
	\hat{G}^\top(t)\right) \hat{x}(t)\,\rd t, \label{eq:xihat} \\
	\hat{\theta}_L(\theta_{NL})&=& \hat{B}^{-1} \int_0^T \hat{G}^\top(t) \left(
	\hat{x}(t) -\hat{\xi} \right) \rd t, \label{eq:thetahat}
\end{eqnarray*}
where $I_d$ denotes the $d \times d$ identity matrix. The notation $\hat{\xi}(\theta_{NL})$ and $\hat{\theta}_L(\theta_{NL})$ emphasizes the dependence of the solution on the nonlinear parameters $\theta_{NL}$. This solution $(\hat{\xi}(\theta_{NL}),\hat{\theta}_L(\theta_{NL}))$ is plugged back into \eqref{eq:int_crit}, yielding the reduced integral criterion function 
\begin{equation}\label{eq:sepnls}
M(\theta_{NL}):=\int_0^T\left|\left|
\hat{x}(t)-\hat{\xi}(\theta_{NL})-\hat{G}(t;\theta_{NL})\hat{\theta}_L(\theta_{NL})
\right|\right|^2\rd t.
\end{equation}
Once $M(\theta_{NL})$ is minimized over $\theta_{NL}$ and a solution 
\begin{eqnarray*}
\hat\theta_{NL}&=&\arg\min_{\theta_{NL}} M(\theta_{NL})
\end{eqnarray*}
 is obtained, estimators for $\xi$ and $\theta$ follow immediately, and are given (with some abuse of the matrix transpose notation) by 
\begin{equation}
\begin{aligned}\label{eq:sls}
\hat\xi_{SLS}&=\hat{\xi}(\hat\theta_{NL}),\\
\hat\theta_{SLS}&=(\hat\theta_{NL},\hat{\theta}_L(\hat\theta_{NL})),
\end{aligned}
\end{equation}
 respectively. Note that the nonlinear optimization is applied only for estimating the nonlinear parameters $\theta_{NL}$, which, in comparison to the NLS approach, can substantially reduce the dimension of the nonlinear optimization problem.

From the above derivation it is clear that SLS problems are a special class of NLS problems, with linear and nonlinear objective functions for different sets of
variables. While the idea of using separability for improving parameter estimation was presented already in \cite{lawton1971elimination}, it seems that much of the subsequent literature is based on the variable projection method proposed by \cite{golub1973differentiation}. \cite{golub2003separable} reviewed 30 years of research into this problem.

\section{Simulation framework and results}\label{sec:exp}

In order to investigate the relative performance of SLS and NLS, we designed and performed a large Monte-Carlo simulation, whose results are presented in this section.

All the computations were done on an Amazon EC2 m5a.4xlarge instance and were carried out in {\bf R} using the \textbf{simode} package of \cite{simode}, which is designed specifically for using separability properties of ODEs. We note in passing that in the context of nonlinear regression, the variable projection method of \cite{golub1973differentiation} is implemented in {\bf R} in the {\bf nls} command; see \cite{ripley2002}, pp.~218--220 for an example of its application; see also the {\bf TIMP} package of \cite{mullen2007timp}. We used default smoothing and optimization settings in {\bf simode}, and in that respect both SLS and NLS received equal treatment; in particular, {\bf simode} uses cross-validation (see, e.g., \cite{wasserman06nonparametric}) to determine the optimal amount of smoothing. The code to reproduce our numerical results can be accessed on GitHub.\footnote{See \href{https://github.com/haroldship/complexity-2019-code/tree/master/Final Code First Submission}{https://github.com/haroldship/complexity-2019-code/tree/master/Final Code First Submission}} For plotting, we relied on the {\bf ggplot2} package in {\bf R}, see \cite{wickham09}.

\subsection{Monte-Carlo study design}
\label{subsec:mc}

We chose several representative and challenging ODE models arising in a variety of scientific disciplines. These were:
\begin{enumerate}[(i)]
	\item SIR for simulating an infection process;
	\item Lotka-Volterra population model with sinusoidal seasonal adjustment;
	\item Generalised Mass Action (GMA) system within BST, e.g., for metabolic pathway systems; 
	\item FitzHugh-Nagumo system of axon potentials.
\end{enumerate}
Further mathematical details on these systems and the specific experimental setups we used are given below.

In each case, we generated observations by numerically integrating the system and next adding independent Gaussian noise to the time courses. We considered various parameter setups, sample sizes, and noise levels, as specified below. The ODE parameters were estimated via both NLS and SLS, defined in equations \eqref{eq:nls} and  \eqref{eq:sls}, respectively. 

As performance criteria, the time required to perform optimization and accuracy of the resulting parameter estimates were used. While comparing computation times is trivial, numerous options are available for comparing accuracy. We focused on the main difference between the two optimization schemes, namely the way they deal with the estimation of linear parameters. SLS does not require initial guesses for these parameters. By contrast, NLS does require a good initial guess for each linear parameter, or otherwise it might diverge or get stuck in a local minimum: finding ``good'' solutions to nonlinear optimization problems requires ``good'' initial guesses in the parameter space. Thus some ``prior information'' regarding these parameters is of crucial importance for optimization purposes. The key insight is that this prior information is encapsulated in the mathematical form of the ODEs themselves, such as \eqref{eq:sep_ode}. Importantly, while NLS does not take into account the special mathematical features of the ODEs and treats all the parameters in a uniform manner, this is not the case for SLS. Thus, one might {\it a priori} expect SLS to be more efficient and possibly more accurate than NLS, when prior information regarding the linear parameters is of low quality. On the other hand, when one has high quality prior information regarding the linear parameters, we expect that SLS and NLS to perform similarly. One might note that the nonlinear parameters in almost all GMA and S-systems are very tightly bounded, usually between -1 and +2, and that their sign is often known, whereas the linear parameters are unbounded in GMA systems and non-negative in S-systems, and nothing is known about their magnitudes (see Chapter 5 of \cite{voit2000computational}). Thus, not needing prior information on the linear parameters in SLS can be a tremendous advantage.

For the Monte-Carlo study, we varied the prior information by using high, medium and low quality initial guesses for the parameter values. Here higher quality means that the initial guesses were closer to the truth. To be more specific, the initial guesses for the linear parameters used by NLS were Gaussian randoms variables centered on the true parameter values and having standard deviations equal to the true parameter multiplied by a prior information value (thus the prior information value can also be understood as the coefficient of variation of the ``prior distribution''). The specific quantification of ``high'', ``medium'' and ``low'' is admittedly somewhat subjective, and varied across the different ODE models, as specified below. For the sake of better and faster convergence of the optimization algorithms (especially NLS), the nonlinear parameters were constrained to a given range, and this range was the same no matter how we varied the prior information on linear parameters. Further, in each Monte-Carlo iteration we used exactly the same (pseudo-random) initial guess for nonlinear parameters for both NLS and SLS. Thus, as far as the information on nonlinear parameters is concerned, this was kept invariant for each benchmark model, irrespective of the prior on linear parameters. Consequently, both algorithms received the same prior information regarding nonlinear parameters, and hence none was treated preferentially.

The noise level (signal-to-noise ratio, SNR) we used is defined as follows. For a given solution $x(t)$ of an ODE equation, we calculate the standard deviation $\sigma_x=\operatorname{std}(x(t_1),...,x(t_n))$. Then SNR of, say, $10\%$ and $20\%$ is given by $\sigma=\sigma_x/10$, and $\sigma=\sigma_x/5$, respectively, where $\sigma$ is the standard deviation of a Gaussian measurement error $\epsilon$ as defined in equation \eqref{eq:obs}. We will refer to these SNRs as ``low noise'' and ``high noise'', respectively (cf.\ \cite{johnstone05}, albeit in a different context). We then compared the mean square errors (MSE) of the resulting parameter estimates, which leads to a valid comparison in statistically identifiable ODE models (see, e.g., \cite{dattner2015} for relevant definitions and results). As another accuracy measure we used the criteria \eqref{eq:int_crit} and \eqref{eq:sepnls} evaluated at optimal parameter values. The two criteria we propose, though reasonable, are different. Hence, they are not expected to be in agreement in every experimental setup. However, the global conclusions reached with them in Section~\ref{sec:con} are coherent and are in favor of SLS. 

We now provide the mathematical details on the models and the experimental setups. 

\subsubsection{Age-group SIR}
The system is an epidemiological model of an SIR-type (Susceptible -- Infected -- Recovered), and includes age group and seasonal components. The epidemic
in each age group $1\leq a \leq M$ and each season $1\leq y \leq L$ is described using two equations for the proportion of susceptible ($S$) and infected ($I$) individuals in the population (the proportion of recovered individuals is given by $1-S-I$):
\begin{equation}\label{eq:sir}
\begin{array}{l}
S_{a,y}^{\prime}(t)=-S_{a,y}(t)\kappa_y\sum_{j=1}^{M}\beta_{a,j}I_{j,y}(t),
\\
I_{a,y}^{\prime}(t)=S_{a,y}(t)\kappa_y\sum_{j=1}^{M}(\beta_{a,j}I_{j,y}(t))-\gamma I_{a,y}(t).
\end{array}
\end{equation}
The parameters of the model are the $M\times M$ transmission matrix $\beta$, the recovery rate $\gamma$, and $\kappa_{2,\ldots,L}$, which signify the relative infectivity of, e.g., influenza virus strains circulating in seasons $2,\ldots,L$ compared to season $1$ ($\kappa_1$ is used as a reference and is fixed at $1$). As shown in \cite{yaarietal18}, taking into account separability characteristics of this model is advantageous for data fitting purposes. Specifically, \eqref{eq:sir} is nonlinear in the initial values $S(0)$, which are typically unknown and have to be estimated. For our purposes it suffices to consider a model with one age group and one season. The following parameter setup was used: $S(0)=0.56, I(0)=1e-04, \beta=6, \gamma=2.3$. We considered two sample sizes, $18 $ and $36$, and two noise levels, $10\%$ and $20\%$. The prior information used  was $\{0.1,0.2,0.3\}$, corresponding to high, medium and low quality, respectively. The size of the Monte-Carlo study was $500$ simulations.

\subsubsection{Lotka-Volterra with seasonal forcing}

As another benchmark we considered an extension of a classical predator-prey model, namely the Lotka-Volterra model including the seasonal forcing of the predation rate, using two additional parameters that control the amplitude ($\epsilon$) and phase ($\omega$) of the forcing:

\begin{equation*}\label{eq:lv_force}
\begin{array}{l}
x_1^{\prime}(t)=\alpha x_1(t)-\beta (1+\epsilon\sin (2\pi (t/T+\omega)))x_1(t)x_2(t),
\\
x_2^{\prime}(t)=\delta (1+\epsilon\sin (2\pi (t/T+\omega)))x_1(t)x_2(t)-\gamma x_2(t).
\end{array}
\end{equation*}
The nonlinear parameters are $\epsilon$ and $\omega$. We considered the dynamics within the time interval $t\in [0,25]$. The parameter setup is given by 
\begin{eqnarray*}
\theta&=&\{\alpha,\beta,\gamma,\delta,\epsilon,\omega\}
\\&=&\{2/3, 4/3, 1.0, 1.0, 0.2, 0.5\},
\end{eqnarray*} 
and initial values are $\{x_1(0),x_2(0)\}=\{0.9, 0.9\}$.  
Four experimental scenarios where studied, corresponding to sample sizes of $100$ and $200$, and SNRs of $10\%$ and $20\%$. The prior information values were $\{0.05,0.1,0.2\}$, corresponding to high, medium and low quality, respectively. The size of the Monte-Carlo study was $500$ simulations.

\subsubsection{GMA system}

The GMA system we analyzed consists of three differential equations in three variables (\cite{voit2000computational}, pp.~84--85). They are:

\begin{equation*}\label{eq:biosimple2}
\begin{array}{l}
x_1^{\prime}(t)=\gamma_{11}x_2^{f_{121}}(t)x_3^{f_{131}}(t)-\gamma_{12}x_1^{f_{112}}x_2^{f_{122}}-\gamma_{13}x_1^{f_{113}}x_3^{f_{133}},
\\
x_2^{\prime}(t)=\gamma_{12}x_1^{f_{112}}x_2^{f_{122}}-\gamma_{22}x_2^{f_{222}},
\\
x_3^{\prime}(t)=\gamma_{13}x_1^{f_{113}}x_3^{f_{133}}-\gamma_{32}x_3^{f_{332}}.
\end{array}
\end{equation*}
Here the linear parameters are the rate constants $\gamma$, while the nonlinear ones are the kinetic orders $f$. Note that the parameters $f$ are allowed to become negative and their sign might not be known too. We considered the dynamics of the system within the time interval $ [0,4]$. The parameter setup is the one presented in \cite{voit2000computational}, namely 
\begin{eqnarray*}
	\theta&=&\{\gamma_{11},f_{121}, f_{131},\gamma_{12},f_{112}, f_{122},\gamma_{13},f_{113},f_{133}, \gamma_{22},f_{222},\gamma_{32},f_{332}\}
	\\&=&\{0.4,-1.0,-1.0,3.0,0.5,-0.1,2.0,0.75,-0.2,1.5,0.5,5.0,0.5\},
\end{eqnarray*} 
and initial values are $\{x_1(0),x_2(0),x_3(0)\}=\{0.5,0.5,1.0\}$.  
Four experimental scenarios were studied: sample sizes of $100$ and $200$, and SNRs of $10\%$ and $20\%$. The prior information values were $\{0.1,0.3,0.5\}$, corresponding to high, medium and low quality, respectively. The size of the Monte-Carlo study was $500$ simulations. Parameter estimation for GMA systems is considered to be a challenging numerical task (\cite{voit2000computational}).

\subsubsection{FitzHugh-Nagumo system}
The FitzHugh-Nagumo system (\cite{fitzhugh1961impulses}, \cite{nagumo1962active}, \cite{fitzhugh69}, \cite{fitzhugh69}) models spike potential activity in a neuron. It is given by
\begin{equation}\label{eq:fn}
\begin{array}{l}
x_1^{\prime}(t)=c\left(x_1(t)-{x_1^{3}(t)}/{3}+x_2(t)\right),
\\
x_2^{\prime}(t)=-({1}/{c})(x_1(t)-a+bx_2(t)).
\end{array}
\end{equation}
This system with two state variables was proposed as a simplification of the model presented in \cite{hodgkin1952quantitative} for
studying and simulating the animal nerve axon. The model is used in neurophysiology as an approximation of the observed spike
potential.

The system \eqref{eq:fn} is linear in parameters $a,$ and  $b$, but nonlinear in $c$. We considered two sample sizes, $n=20$ and $n=40$, and two SNRs of $10\%$ and $20\%$. The parameters were set to $\{a,b,c\}=\{0.2,0.2,3\}$. The initial values were $\{x_1(0),x_2(0)\}=\{-1.0,1.0\}$. The true solutions were obtained over the time interval $[0,20]$. The prior information used here was $\{0.5,1.0,3.0\}$, corresponding to high, medium and low quality, respectively.\footnote{The initial guesses for parameters were assured to be positive.} The size of the Monte-Carlo study was $500$ simulations. Many researchers studied the problem of parameter estimation for the FitzHugh-Nagumo model. In particular, \cite{ramsay2007parameter}, \cite{campbell2012smooth} and \cite{ramsay17} pointed 
out several difficulties in estimating the parameters for this ODE system.

\subsection{Results of the Monte-Carlo analysis}\label{subsec:results}
Our findings are presented through charts and tables. The primary summaries are Tables \ref{table:linearsmall} and \ref{table:linearlarge}, where we report the ratios of the mean square errors (square errors averaged over Monte-Carlo simulations) for estimates of linear parameters (for nonlinear parameters, see the discussion at the end of this section). Several conclusions can be gleaned from the tables.
\begin{enumerate}[(i)]
	\item Given high-quality prior information, the accuracy of NLS and SLS is comparable, and neither method has a clear lead throughout the variety of experimental setups.\footnote{At least some of the differences that one sees from the raw numbers in the tables are plausibly attributable to the Monte-Carlo simulation error and as such appear to be insignificant.}
	\item When the quality of prior information degrades to medium or low, the performance of SLS becomes in most cases significantly better than that of NLS (with an extent depending on the specific experimental setup).
	\item For a fixed noise level, as the sample size increases, the advantage of SLS becomes more pronounced.
	\item For a fixed sample size, as the noise level increases, the SLS is still better than NLS, but to a lesser extent.
\end{enumerate}

{\small
	\begin{table}
		\begin{center}
			\captionsetup{width=0.85\textwidth}
			\caption{MSE ratios for linear parameters (small samples)}
			\begin{tabular}{l@{\hskip 0.135in}lrcl@{\hskip 0.33in}lrcl@{}} %
				\toprule
				& \multicolumn{4}{c}{{\bf \hskip -0.33in Low noise}} & \multicolumn{4}{c}{{\bf High noise}}\\
				\cmidrule(l{0.1in}r{0.43in}){2-5}  \cmidrule(r{0.15in}l{0.1in}){6-9}
				{\bf Prior} & {{\bf{sir}} } & {{\bf{ltk}}} & \multicolumn{1}{c}{{\bf{gma}}} & \multicolumn{1}{l}{{ \hskip -0.05in \bf{ftz}}}
				& {{\bf{sir}} } & {{\bf{ltk}}} & \multicolumn{1}{c}{{\bf{gma}}} & \multicolumn{1}{l}{{ \hskip -0.05in \bf{ftz}}}  \\
				\midrule
				{\bf low}     & 8.3 & 5.8 & 4.0 & 2.3    & 2.8 & 2.2  & 2.2  & 1.3  \\
				{\bf medium}  & 3.9 & 1.8 & 3.1 & 1.7    & 1.4 & 1.2  & 1.8  & 1.2  \\
				{\bf high}    & 0.9 & 1.3 & 0.9 & 2.0    & 0.4 & 1.0  & 0.6  & 1.1  \\
				\bottomrule
			\end{tabular}
			\caption*{
				{\small
					
					The table gives the MSE ratios (computed by averaging square errors over Monte-Carlo simulation runs) of NLS and SLS for estimating the linear parameters in various benchmark models and under different experimental setups (see Section \ref{subsec:mc} for detailed specifications). To identify model names, self-explanatory abbreviations are used. The values in the table are rounded off to one digit after zero. The sample size is $n=100$ for the GMA and Lotka-Volterra models, $n=20$ for the FitzHugh-Nagumo system, and  $n=18$ for SIR model. The noise levels are $10\%$ and $20\%.$ Values larger than $1$ in the table correspond to the cases where SLS performs better than NLS. Note the decreasing pattern in the columns, reflecting the effect of the quality of prior information on the performance of NLS.}}
			\label{table:linearsmall}
		\end{center}
	\end{table}
}

{\small
	\begin{table}
		\begin{center}
			\captionsetup{width=0.85\textwidth}
			\caption{MSE ratios for linear parameters (large samples)}
				\begin{tabular}{l@{\hskip 0.135in}r@{\hskip 0.135in}rcl@{\hskip 0.33in}lrcl@{}} %
					\toprule
					& \multicolumn{4}{c}{{\bf \hskip -0.33in Low noise}} & \multicolumn{4}{c}{{\bf High noise}}\\
					\cmidrule(l{0.13in}r{0.44in}){2-5}  \cmidrule(r{0.15in}l{0.1in}){6-9}
					{\bf Prior} & \multicolumn{1}{l}{{\hskip 0.01in \bf{sir}} } & {{\bf{ltk}}} & \multicolumn{1}{c}{{\bf{gma}}} & \multicolumn{1}{l}{{ \hskip -0.05in \bf{ftz}}}
					& {{\bf{sir}} } & {{\bf{ltk}}} & \multicolumn{1}{c}{{\bf{gma}}} & \multicolumn{1}{l}{{ \hskip -0.05in \bf{ftz}}}  \\
				\midrule
				{\bf low}   & 12.0 & 9.6 & 3.9 & 3.9     & 3.8 & 3.2 & 2.2 & 2.2 \\
				{\bf medium}& 4.1  & 4.3 & 2.8 & 1.9     & 1.6 & 1.6 & 1.6 & 1.3  \\
				{\bf high}  & 1.0  & 2.2 & 0.7 & 2.3     & 0.7 & 1.2 & 0.5 & 1.5  \\
				\bottomrule
			\end{tabular}
			\caption*{
				{\small
					
				The sample size is $n=200$ for the GMA and Lotka-Volterra models; $n=40$ for the FitzHugh-Nagumo system; and $n=36$ for the SIR model. The noise levels are $10\%$ and $20\%.$ For an interpretation of the results, see Table \ref{table:linearsmall}. Note an increased advantage of SLS over NLS in comparison to Table  \ref{table:linearsmall}.}}
			\label{table:linearlarge}
		\end{center}
	\end{table}
}

\begin{figure}
	\centering
	\captionsetup{width=0.85\textwidth}
	\includegraphics[width=0.85\textwidth]{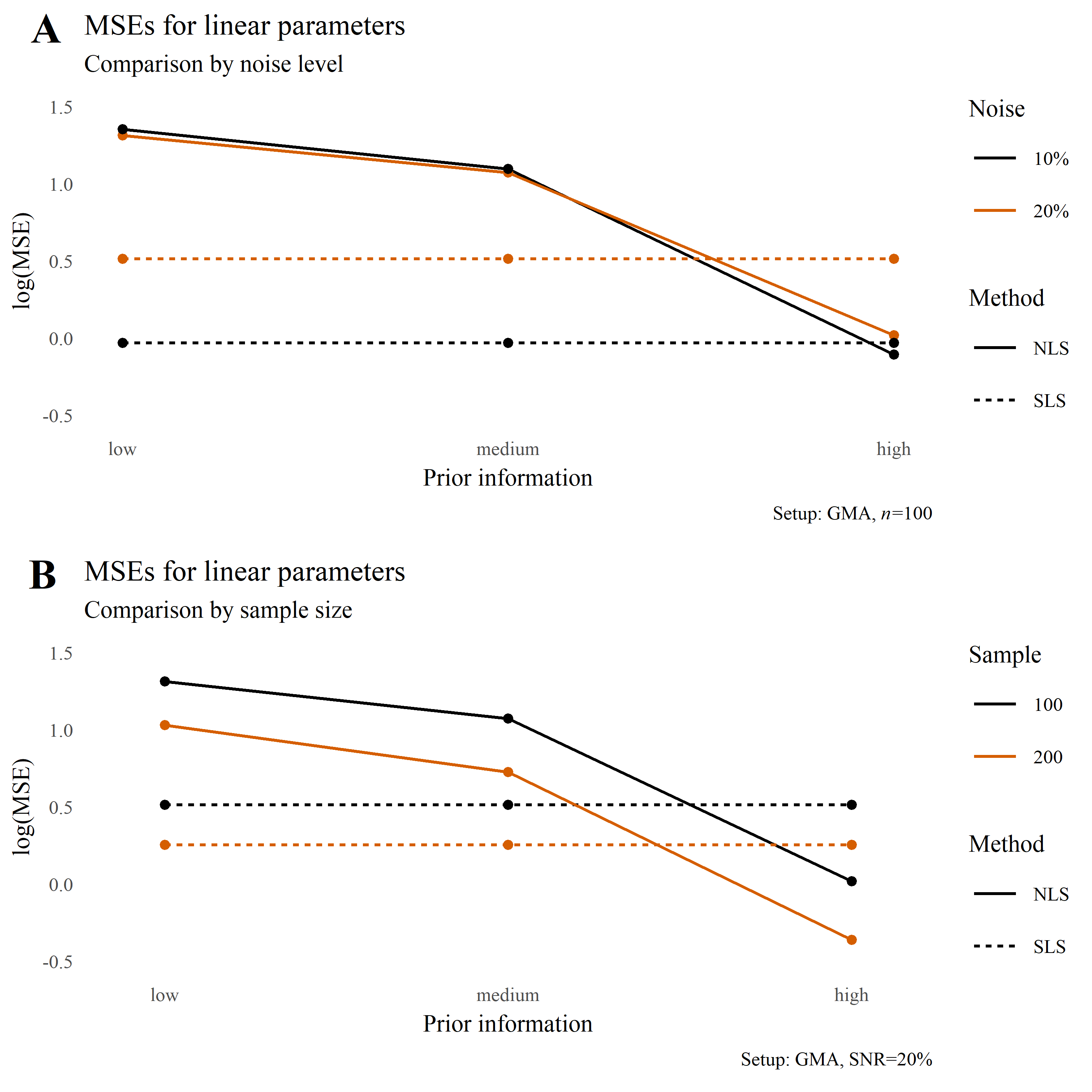}
	\caption{\label{fig:mse} {\small The plot gives MSEs on a log scale (computed as averages over Monte-Carlo simulation runs) for linear parameters plotted against the quality of prior information. In the top panel, labeled A, the comparison is on the basis of the noise level. The graph indicates that the performance of NLS worsens with lowering of the quality of prior information. On the other hand, the performance of SLS is not affected by the quality of prior information, in agreement with the experimental design. Except for the rare case of high quality prior information, where NLS is better, SLS clearly outperforms NLS. In the bottom panel, labeled B, the comparison is based on the sample size.	The overall pattern is similar to that in the top panel.} 
	}
\end{figure}

These points can also be visualized through a combination of simple statistical charts. Thus, Figure \ref{fig:mse} displays the line graphs that compare MSEs of the two methods under several experimental setups. Whereas the numbers in Tables \ref{table:linearsmall} and \ref{table:linearlarge} are ratios of MSEs, in the figures we present the absolute MSE values. From the graphs, an advantage of SLS over NLS is apparent for less than ideal prior information. Note that in this specific setting SLS performed worse than NLS for high quality prior information. A plausible explanation lies in the fact that while under our experimental setup the amount of information used by SLS via \eqref{eq:sep_ode} is fixed throughout simulations, NLS can in principle receive arbitrarily precise initial guesses on linear parameters. One may therefore envision existence of a point, from where on using the latter kind of information outweighs the benefits of using the structural relationship \eqref{eq:sep_ode}. However, a precise quantification of the phenomenon is hardly possible beyond an observation that it appears to manifest itself in scenarios with excellent knowledge on likely parameter values. In reality, ideal prior information is rare. 

The bottom panel of Figure \ref{fig:mse} further suggests that in the specific scenarios that we report there, SLS improves when the noise level diminishes; this is unlike NLS in that figure.

\begin{figure}
	\centering
	\captionsetup{width=0.85\textwidth}
	\includegraphics[trim=0cm 1.8cm 0cm 1.8cm, clip,width=0.85\textwidth]{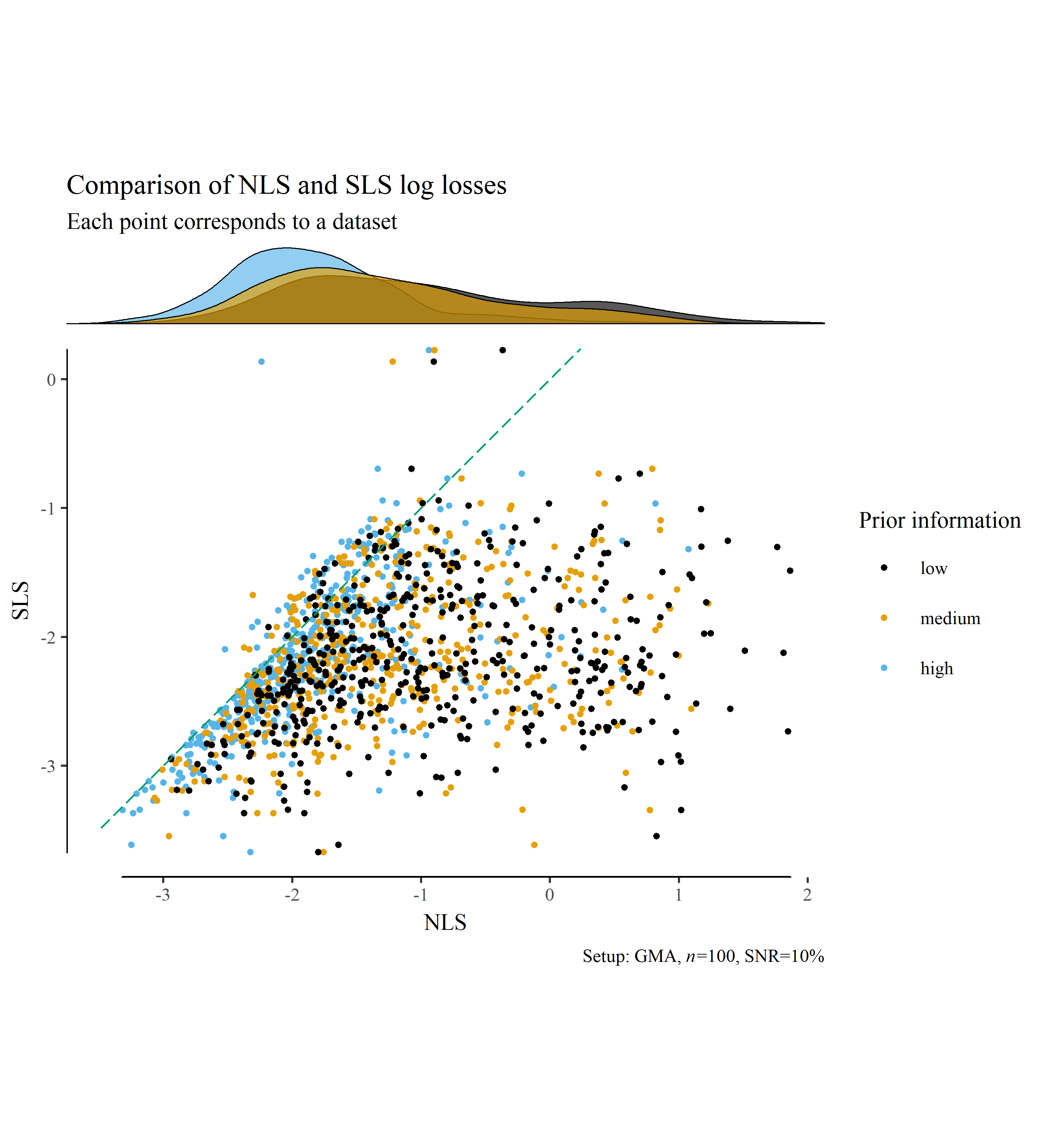}
	\caption{\label{fig:loss_scatter} {\small The plot visualizes the performance (on a log scale) of NLS and SLS according to criteria \eqref{eq:int_crit} and \eqref{eq:sepnls}, which are evaluated at the optimal parameter estimates. Points in the scatterplot are colored according to the quality of prior information used to compute the NLS estimates. The 45\textdegree{} diagonal line passing through the origin has been added for reference and intuitive assessment. The scatterplot is supplemented with marginal density estimates using the same color coding. The density estimates indicate that, as the quality of prior information degrades, the quality of NLS results suffers, which manifests in longer right tails of the densities. By definition, performance of SLS is not affected by the quality of prior information on linear parameters. For high quality prior information, clustering of losses in the scatterplot close to the reference line suggests that the overall performance of both NLS and SLS is comparable. As the quality of prior information decreases, the point clouds spread to the right, indicating that SLS starts to perform noticeably better than NLS. Furthermore, unlike Tables \ref{table:linearsmall} and \ref{table:linearlarge}, the scatterplot and the range frame (see \cite{tufte01}, pp.~130--132) convey an impression of the variability in the estimation results over multiple datasets: NLS is visually more variable than SLS.}
	}
\end{figure}

Figure \ref{fig:loss_scatter} is a scatterplot of NLS and SLS losses \eqref{eq:int_crit} and \eqref{eq:sepnls} (on a log scale) evaluated at optimal parameter estimates. The figure highlights in yet another way the importance of prior information for NLS: it is evident that the performance of the latter is strongly affected by the quality of initial parameter guesses. Again, NLS and SLS perform similarly when the prior information is of high quality. However, when the quality of prior information is less than ideal, as it is in most applications, NLS becomes substantially worse than SLS. The scatterplot also gives a quick impression of the variability of estimation results.

\begin{figure}
	\centering
	\captionsetup{width=0.85\textwidth}
	\includegraphics[width=0.85\textwidth]{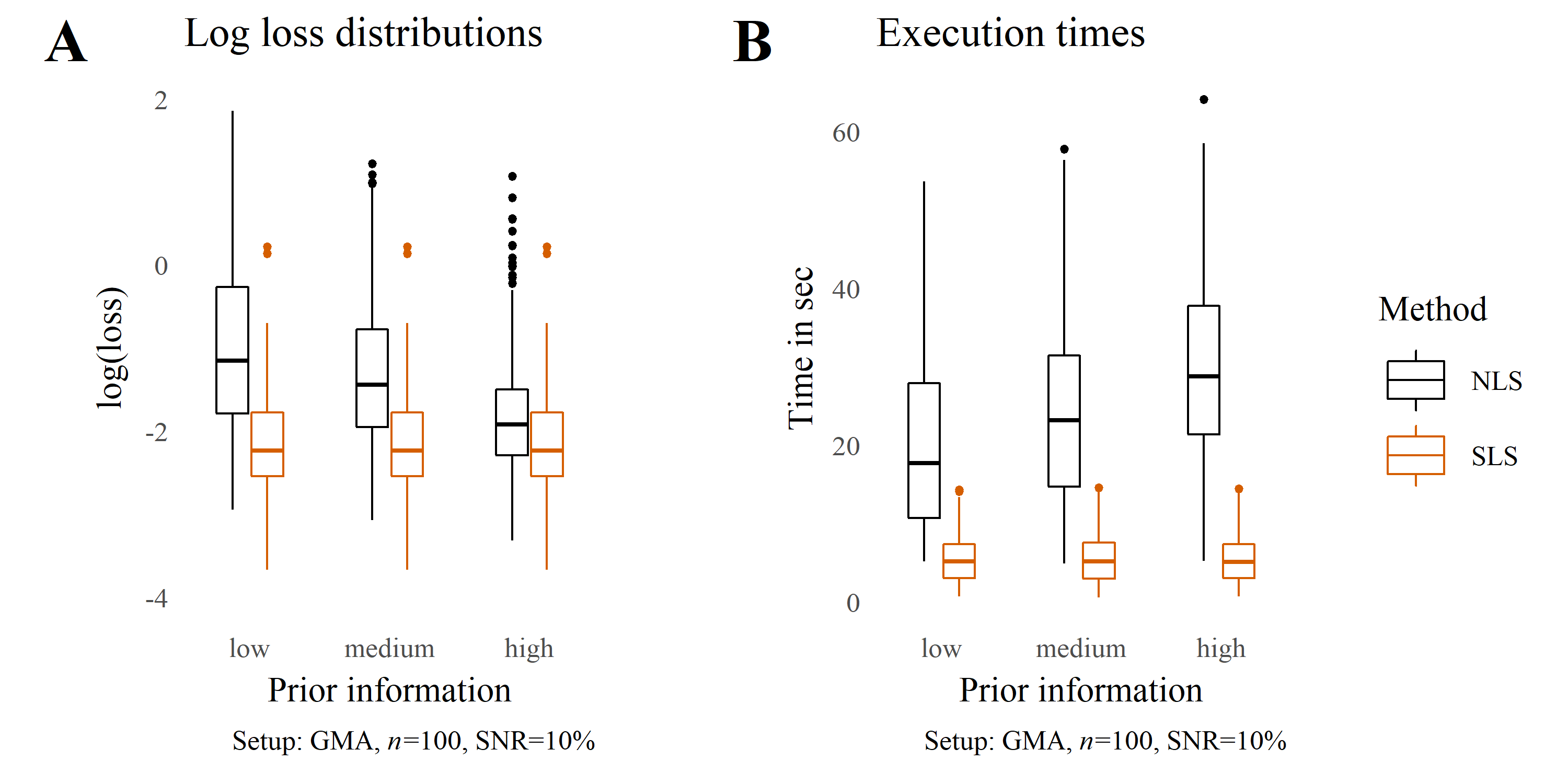}
	\caption{\label{fig:bp_loss} {\small The plot presents a comparison of NLS and SLS. In the left panel, labeled A, boxplots of the losses \eqref{eq:int_crit} and \eqref{eq:sepnls} (on a log scale) evaluated at the optimal parameter estimates are displayed. For high quality prior information, the NLS and SLS loss distributions are close. As the quality of prior information degrades, NLS losses start to take higher values compared to SLS, and their variability increases, as evidenced by the elongation of boxplots. In the right panel, labeled B, the computation times are compared. The NLS computation times tend to be longer than those of SLS, and increase as the quality of prior information increases. In both panels, the performance of SLS does not vary with the quality of prior information, in concordance with the experimental design. }
	}
\end{figure}

The conclusions that we drew from Figure \ref{fig:loss_scatter} are confirmed by the left panel of Figure \ref{fig:bp_loss}, which presents boxplots of NLS and SLS losses (on a log scale) measured according to criteria \eqref{eq:int_crit} and \eqref{eq:sepnls}. The pattern is clear: SLS is better than NLS, and the inferiority for NLS becomes more dramatic with degrading prior information.

The right panel of Figure \ref{fig:bp_loss} summarizes computation times. SLS is in general much faster. The execution time of NLS is affected by the quality of prior information, and interestingly, increases with this quality.

{\small
	\begin{table}
		\begin{center}
			\captionsetup{width=0.85\textwidth}
			\caption{MSE ratios for nonlinear parameters (small samples)}
			\begin{tabular}{l@{\hskip 0.135in}r@{\hskip 0.135in}rcl@{\hskip 0.33in}lrcl@{}} %
				\toprule
				& \multicolumn{4}{c}{{\bf \hskip -0.33in Low noise}} & \multicolumn{4}{c}{{\bf High noise}}\\
				\cmidrule(l{0.13in}r{0.44in}){2-5}  \cmidrule(r{0.15in}l{0.1in}){6-9}
				{\bf Prior} & \multicolumn{1}{l}{{\hskip 0.01in \bf{sir}} } & {{\bf{ltk}}} & \multicolumn{1}{c}{{\bf{gma}}} & \multicolumn{1}{l}{{ \hskip -0.05in \bf{ftz}}}
				& {{\bf{sir}} } & {{\bf{ltk}}} & \multicolumn{1}{c}{{\bf{gma}}} & \multicolumn{1}{l}{{ \hskip -0.05in \bf{ftz}}}  \\
				\midrule
				{\bf low}    & 23.0 & 1.7 & 1.1 & 1.1    & 6.6 & 1.1 & 1.0 & 1.0  \\
				{\bf medium} & 8.7  & 1.2 & 1.0 & 1.0    & 2.6 & 0.9 & 1.0 & 1.0 \\
				{\bf high}   & 1.0  & 1.0 & 0.9 & 1.0    & 0.4 & 0.9 & 0.9 & 1.0 \\
				\bottomrule
			\end{tabular}
			\caption*{
				{\small
					
					The table gives the MSE ratios (computed through square errors averaged over Monte-Carlo simulations) of NLS and SLS for estimating the nonlinear parameters. The experimental setup is as in Table \ref{table:linearsmall}. Values larger than $1$ in the table correspond to the cases where SLS performs better than NLS.
					Since the prior information regarding nonlinear parameters stays invariant (see Section \ref{subsec:mc} for details), the table in particular shows the effects that the quality of initial guesses for linear parameters has on the estimation accuracy of NLS in the case of nonlinear ones. The results suggest that in some settings a vague prior knowledge on linear parameters may have an adversary effect on the accuracy of NLS for nonlinear ones too.
			}}
			\label{table:nonlinearsmall}
		\end{center}
	\end{table}
}

{\small
	\begin{table}
		\begin{center}
			\captionsetup{width=0.85\textwidth}
			\caption{MSE ratios for nonlinear parameters (large samples)}
			\begin{tabular}{l@{\hskip 0.135in}r@{\hskip 0.135in}rcl@{\hskip 0.33in}lrcl@{}} %
				\toprule
				& \multicolumn{4}{c}{{\bf \hskip -0.33in Low noise}} & \multicolumn{4}{c}{{\bf High noise}}\\
				\cmidrule(l{0.13in}r{0.44in}){2-5}  \cmidrule(r{0.15in}l{0.1in}){6-9}
				{\bf Prior} & \multicolumn{1}{l}{{\hskip 0.01in \bf{sir}} } & {{\bf{ltk}}} & \multicolumn{1}{c}{{\bf{gma}}} & \multicolumn{1}{l}{{ \hskip -0.05in \bf{ftz}}}
				& {{\bf{sir}} } & {{\bf{ltk}}} & \multicolumn{1}{c}{{\bf{gma}}} & \multicolumn{1}{l}{{ \hskip -0.05in \bf{ftz}}}  \\
				\midrule
				{\bf low}     & 29.0 & 4.1 & 1.1 & 1.6    & 6.5 & 1.3 & 1.0 & 1.3 \\
				{\bf medium}  & 8.2  & 2.1 & 0.9 & 1.0    & 2.2 & 1.1 & 0.9 & 1.0 \\
				{\bf high}    & 0.9  & 1.2 & 0.8 & 1.0    & 0.7 & 0.9 & 0.8 & 1.0 \\
				\bottomrule
			\end{tabular}
			\caption*{
				{\small
					
					The setup is as in Table \ref{table:linearlarge}. For an interpretation of the results see Table \ref{table:nonlinearsmall}.
			}}
			\label{table:nonlinearlarge}
		\end{center}
	\end{table}
}

Finally, in Tables \ref{table:nonlinearsmall} and \ref{table:nonlinearlarge} we provide information regarding the nonlinear parameters, where in the case of NLS one can observe how the prior knowledge on linear parameters propagates itself into estimation accuracy for nonlinear ones. In particular, for less than ideal prior information on the linear parameters, SLS holds a pronounced edge over NLS also in the case of nonlinear parameters.

\section{Conclusions and outlook}\label{sec:con}
In this work, we designed an extensive simulation study to explore the relative statistical and computational performance of two optimization schemes for inference in dynamic systems: the typical nonlinear least-squares (NLS) method and a new separable nonlinear least-squares (SLS) approach. As benchmarks, we considered several widely used ODE models arising in a variety of scientific fields. We measured statistical performance of the two methods by the mean square error (MSE) of the estimates. As another performance criterion, we employed the loss function values at the optimal parameter estimates. Computational performance of the methods was also compared by the execution times required to complete each optimization.

A pattern that emerged from our study is that SLS in general performs at least as well as, and frequently better than NLS, especially if the prior information on optimal parameters is not ideal, which is typically the case in practice. This statement is uniformly true over all models tested. 

Our recommendation therefore is that parameter estimation problems for complex dynamic systems should be addressed, whenever the system contains an appreciable number of linear parameters, with the separable nonlinear least-squares method, rather than the more commonly used nonlinear least-squares method. 

While the above message is simple and unambiguous, one must stress that data fitting in complex dynamical systems remains a challenging problem that cannot be treated in a cavalier fashion, even if one takes advantage of separability. For instance, in order to uncover the patterns in Section \ref{sec:exp} of this work, we had to carefully design the experimental study, because otherwise simulations might not have converged, or might have converged to poor solutions. This was true for both NLS and SLS, but whenever they were observed, convergence issues were much more severe for NLS (especially sensitive was the case of the FitzHugh-Nagumo system). This result highlights the crucial role of {\it prior information} regarding the parameters, or expressed differently, the quality of the initial parameter guesses used for the optimization. We focused here primarily on the effects of the prior information on the linear parameters. However, it also became clear that prior information on the nonlinear parameters has an equally crucial role for optimization purposes, this being true for both NLS and SLS (data not shown). 

As a result of our exploratory work, we envision the following promising research directions for the future.

\begin{enumerate}[(i)]
	\item \textbf{Numerical implementation of SLS for dynamic systems}. All the computations in our paper were done in {\bf R} using the \textbf{simode} package of \cite{simode}. However, the idea of using separability properties of ODEs is independent of a particular programming language and can be implemented within other software packages quite as well. Indeed, much work has been done in the context of the variable projection method since it was first introduced in \cite{golub1973differentiation}. We are aware of  the {\bf TIMP} package of \cite{mullen2007timp}, which implements the variable projection method. Thus, a next step could be to combine the strengths of both packages, \textbf{simode} and {\bf TIMP}, in order to develop advanced software for variable projection in the context of dynamic systems. 
	\item \textbf{Customized algorithms for specific classes of complex dynamical systems}. It is well-known that the performance of an optimization scheme depends crucially on the underlying mathematical model used for description of the data. Thus, it appears that different classes of dynamic models require specific algorithms tailored to their peculiarities. For instance, parameter estimation for GMA systems has different challenges than those encountered when working with SIR (see Section \ref{sec:exp}). We expect that there is much to gain from focusing future research on specific classes of models and developing stable algorithms for their parameter estimation. 
	\item \textbf{Theoretical properties of SLS in the context of dynamic systems}. \cite{gugushvili2012sqrt}  studied the statistical properties of NLS in the general context of smoothing, while \cite{dattner2015} specifically addressed ODE systems that are linear in (functions of) the parameters. One might expect that some assumptions used in \cite{gugushvili2012sqrt} can be relaxed when the problem is closer to the one considered in \cite{dattner2015}. 
	\item \textbf{Extensions to partially observed, high-dimensional, and misspecified dynamic systems}. Recent work dealing with inference in high-dimensional ODE models suggests that exploiting linearity in parameters is crucial for developing a successful estimation methodology (see, e.g., \cite{chen2017network} and \cite{wu2018parameter}). More generally, it would be interesting to study with the variable projection method the cases of partially observed, high-dimensional, and possibly misspecified dynamic systems. This might additionally require the use of high-dimensional regularization techniques (e.g., \cite{chen2017network}) for balancing data and model, and specifically taking into account a potential model misspecification (see \cite{ramsay2007parameter}).  
	\end{enumerate}


\bibliographystyle{apa-good}
\bibliography{refs}

\begin{thebibliography}{53}
\expandafter\ifx\csname natexlab\endcsname\relax\def\natexlab#1{#1}\fi
\expandafter\ifx\csname url\endcsname\relax
  \def\url#1{{\tt #1}}\fi
\expandafter\ifx\csname urlprefix\endcsname\relax\def\urlprefix{URL }\fi

\bibitem[{Basu \& Bresler(2000)}]{basu2000stability}
Basu, S., \& Bresler, Y. (2000).
\newblock {The stability of nonlinear least squares problems and the
  Cram{\'e}r-Rao bound}.
\newblock {\em IEEE Transactions on Signal Processing\/}, {\em 48\/}(12),
  3426--3436.

\bibitem[{Brunel(2008)}]{brunel2008parameter}
Brunel, N. J.~B. (2008).
\newblock {Parameter estimation of ODE's via nonparametric estimators}.
\newblock {\em Electron. J. Stat.\/}, {\em 2\/}, 1242--1267.

\bibitem[{Campbell \& Steele(2012)}]{campbell2012smooth}
Campbell, D., \& Steele, R.~J. (2012).
\newblock Smooth functional tempering for nonlinear differential equation
  models.
\newblock {\em Stat. Comput.\/}, {\em 22\/}(2), 429--443.

\bibitem[{Chen et~al.(2017)Chen, Shojaie, \& Witten}]{chen2017network}
Chen, S., Shojaie, A., \& Witten, D.~M. (2017).
\newblock Network reconstruction from high-dimensional ordinary differential
  equations.
\newblock {\em Journal of the American Statistical Association\/}, {\em
  112\/}(520), 1697--1707.

\bibitem[{Chou \& Voit(2009)}]{chou2009recent}
Chou, I.-C., \& Voit, E.~O. (2009).
\newblock Recent developments in parameter estimation and structure
  identification of biochemical and genomic systems.
\newblock {\em Math. Biosci.\/}, {\em 219\/}(2), 57--83.

\bibitem[{Chung \& Nagy(2010)}]{chung2010efficient}
Chung, J., \& Nagy, J.~G. (2010).
\newblock An efficient iterative approach for large-scale separable nonlinear
  inverse problems.
\newblock {\em SIAM J. Sci. Comput.\/}, {\em 31\/}(6), 4654--4674.

\bibitem[{Dattner(2015)}]{dattner2015model}
Dattner, I. (2015).
\newblock A model-based initial guess for estimating parameters in systems of
  ordinary differential equations.
\newblock {\em Biometrics\/}, {\em 71\/}(4), 1176--1184.

\bibitem[{Dattner \& Gugushvili(2018)}]{dattnergugushvili18}
Dattner, I., \& Gugushvili, S. (2018).
\newblock Application of one-step method to parameter estimation in {ODE}
  models.
\newblock {\em Stat. Neerl.\/}, {\em 72\/}(2), 126--156.

\bibitem[{Dattner \& Huppert(2018)}]{dattner2018modern}
Dattner, I., \& Huppert, A. (2018).
\newblock Modern statistical tools for inference and prediction of infectious
  diseases using mathematical models.
\newblock {\em Stat. Methods Med. Res.\/}, {\em 27\/}(7), 1927--1929.

\bibitem[{Dattner \& Klaassen(2015)}]{dattner2015}
Dattner, I., \& Klaassen, C. A.~J. (2015).
\newblock Optimal rate of direct estimators in systems of ordinary differential
  equations linear in functions of the parameters.
\newblock {\em Electron. J. Statist.\/}, {\em 9\/}(2), 1939--1973.

\bibitem[{Dattner et~al.(2017)Dattner, Miller, Petrenko, Kadouri, Jurkevitch,
  \& Huppert}]{dattner2017modelling}
Dattner, I., Miller, E., Petrenko, M., Kadouri, D.~E., Jurkevitch, E., \&
  Huppert, A. (2017).
\newblock Modelling and parameter inference of predator--prey dynamics in
  heterogeneous environments using the direct integral approach.
\newblock {\em Journal of The Royal Society Interface\/}, {\em 14\/}(126),
  20160525.

\bibitem[{{Erichson} et~al.(2018){Erichson}, {Zheng}, {Manohar}, {Brunton},
  {Kutz}, \& {Aravkin}}]{erichson2018sparse}
{Erichson}, N.~B., {Zheng}, P., {Manohar}, K., {Brunton}, S.~L., {Kutz}, J.~N.,
  \& {Aravkin}, A.~Y. (2018).
\newblock {Sparse principal component analysis via variable projection}.
\newblock {\em arXiv e-prints\/}.
\newline\urlprefix\url{https://arxiv.org/abs/1804.00341}

\bibitem[{{Fan} \& {Gijbels}(1996)}]{fan1996}
{Fan}, J., \& {Gijbels}, I. (1996).
\newblock {\em {Local polynomial modelling and its applications}\/}, vol.~66 of
  {\em {Monographs on Statistics and Applied Probability}\/}.
\newblock London: Chapman \& Hall.

\bibitem[{FitzHugh(1961)}]{fitzhugh1961impulses}
FitzHugh, R. (1961).
\newblock Impulses and physiological states in theoretical models of nerve
  membrane.
\newblock {\em Biophysical Journal\/}, {\em 1\/}(6), 445--466.

\bibitem[{FitzHugh(1969)}]{fitzhugh69}
FitzHugh, R. (1969).
\newblock Mathematical models of excitation and propagation in nerve.
\newblock In H.~P. Schwan (Ed.) {\em Biological engineering\/}, vol.~9 of {\em
  Inter-university electronics series\/}, chap.~1, (pp. 1--85). New York, NY:
  McGraw-Hill.

\bibitem[{{Gan} et~al.(2018){Gan}, {Chen}, {Chen}, \& {Chen}}]{gan2017some}
{Gan}, M., {Chen}, C. L.~P., {Chen}, G., \& {Chen}, L. (2018).
\newblock On some separated algorithms for separable nonlinear least squares
  problems.
\newblock {\em IEEE Transactions on Cybernetics\/}, {\em 48\/}(10), 2866--2874.

\bibitem[{{Gennemark} \& {Wedelin}(2007)}]{gennemark2007}
{Gennemark}, P., \& {Wedelin}, D. (2007).
\newblock Efficient algorithms for ordinary differential equation model
  identification of biological systems.
\newblock {\em IET Systems Biology\/}, {\em 1\/}(2), 120--129.

\bibitem[{{Girolami} \& {Calderhead}(2011)}]{zbMATH07045664}
{Girolami}, M., \& {Calderhead}, B. (2011).
\newblock {Riemann manifold Langevin and Hamiltonian Monte Carlo methods. With
  discussion and authors' reply}.
\newblock {\em {J. R. Stat. Soc., Ser. B, Stat. Methodol.}\/}, {\em 73\/}(2),
  123--214.

\bibitem[{Golub \& Pereyra(2003)}]{golub2003separable}
Golub, G., \& Pereyra, V. (2003).
\newblock {Separable nonlinear least squares: The variable projection method
  and its applications}.
\newblock {\em {Inverse Probl.}\/}, {\em 19\/}(2), 1--26.

\bibitem[{Golub \& Pereyra(1973)}]{golub1973differentiation}
Golub, G.~H., \& Pereyra, V. (1973).
\newblock The differentiation of pseudo-inverses and nonlinear least squares
  problems whose variables separate.
\newblock {\em SIAM J. Numer. Anal.\/}, {\em 10\/}(2), 413--432.

\bibitem[{{Green} \& {Silverman}(1994)}]{green1994}
{Green}, P.~J., \& {Silverman}, B.~W. (1994).
\newblock {\em {Nonparametric regression and generalized linear models: a
  roughness penalty approach}\/}, vol.~58 of {\em {Monographs on Statistics and
  Applied Probability}\/}.
\newblock London: Chapman \& Hall.

\bibitem[{Gugushvili \& Klaassen(2012)}]{gugushvili2012sqrt}
Gugushvili, S., \& Klaassen, C. A.~J. (2012).
\newblock {\(\sqrt{n}\)-consistent parameter estimation for systems of ordinary
  differential equations: bypassing numerical integration via smoothing}.
\newblock {\em {Bernoulli}\/}, {\em 18\/}(3), 1061--1098.

\bibitem[{Himmelblau et~al.(1967)Himmelblau, Jones, \&
  Bischoff}]{himmelblau1967determination}
Himmelblau, D., Jones, C., \& Bischoff, K. (1967).
\newblock Determination of rate constants for complex kinetics models.
\newblock {\em Industrial \& Engineering Chemistry Fundamentals\/}, {\em
  6\/}(4), 539--543.

\bibitem[{Hodgkin \& Huxley(1952)}]{hodgkin1952quantitative}
Hodgkin, A.~L., \& Huxley, A.~F. (1952).
\newblock A quantitative description of membrane current and its application to
  conduction and excitation in nerve.
\newblock {\em The Journal of Physiology\/}, {\em 117\/}(4), 500--544.

\bibitem[{{Johnstone} \& {Silverman}(2005)}]{johnstone05}
{Johnstone}, I.~M., \& {Silverman}, B.~W. (2005).
\newblock {Empirical Bayes selection of wavelet thresholds}.
\newblock {\em {Ann. Stat.}\/}, {\em 33\/}(4), 1700--1752.

\bibitem[{Lawton \& Sylvestre(1971)}]{lawton1971elimination}
Lawton, W.~H., \& Sylvestre, E.~A. (1971).
\newblock Elimination of linear parameters in nonlinear regression.
\newblock {\em Technometrics\/}, {\em 13\/}(3), 461--467.

\bibitem[{{Lotka}(1956)}]{lotka1956}
{Lotka}, A.~J. (1956).
\newblock {\em {Elements of mathematical biology}\/}.
\newblock {New York}: Dover Publications, Inc.
\newblock {Unabridged republication of the first edition published under the
  title: Elements of physical biology}.

\bibitem[{{May}(2001)}]{may2001}
{May}, R.~M. (2001).
\newblock {\em {Stability and complexity in model ecosystems. With a new
  introduction by the author}\/}.
\newblock Princeton, NJ: Princeton University Press, 2nd ed.

\bibitem[{McGoff et~al.(2015)McGoff, Mukherjee, \& Pillai}]{mmp2012dsreview}
McGoff, K., Mukherjee, S., \& Pillai, N. (2015).
\newblock {Statistical inference for dynamical systems: A review}.
\newblock {\em Statist. Surv.\/}, {\em 9\/}, 209--252.

\bibitem[{Mullen \& van Stokkum(2007)}]{mullen2007timp}
Mullen, K., \& van Stokkum, I. (2007).
\newblock {TIMP: An R package for modeling multi-way spectroscopic
  measurements}.
\newblock {\em Journal of Statistical Software\/}, {\em 18\/}(3), 1--46.

\bibitem[{Mullen(2008)}]{mullen2008separable}
Mullen, K.~M. (2008).
\newblock {\em Separable nonlinear models: theory, implementation and
  applications in physics and chemistry\/}.
\newblock Phd thesis, Vrije Universiteit.

\bibitem[{Nagumo et~al.(1962)Nagumo, Arimoto, \& Yoshizawa}]{nagumo1962active}
Nagumo, J., Arimoto, S., \& Yoshizawa, S. (1962).
\newblock An active pulse transmission line simulating nerve axon.
\newblock {\em Proceedings of the IRE\/}, {\em 50\/}(10), 2061--2070.

\bibitem[{Peschel \& Mende(1986)}]{peschel1986predator}
Peschel, M., \& Mende, W. (1986).
\newblock {\em The predator-prey model: do we live in a Volterra world?\/}.
\newblock Springer.

\bibitem[{{Ramsay} \& {Hooker}(2017)}]{ramsay17}
{Ramsay}, J., \& {Hooker}, G. (2017).
\newblock {\em {Dynamic data analysis. Modeling data with differential
  equations}\/}.
\newblock {Springer Series in Statistics}. New York, NY: Springer.

\bibitem[{Ramsay et~al.(2007)Ramsay, Hooker, Campbell, \&
  Cao}]{ramsay2007parameter}
Ramsay, J.~O., Hooker, G., Campbell, D., \& Cao, J. (2007).
\newblock Parameter estimation for differential equations: a generalized
  smoothing approach.
\newblock {\em {J. R. Stat. Soc., Ser. B, Stat. Methodol.}\/}, {\em 69\/}(5),
  741--796.

\bibitem[{Ruhe \& Wedin(1980)}]{ruhe1980algorithms}
Ruhe, A., \& Wedin, P.~{\AA}. (1980).
\newblock Algorithms for separable nonlinear least squares problems.
\newblock {\em SIAM Review\/}, {\em 22\/}(3), 318--337.

\bibitem[{{Savageau}(1976)}]{savageau1976}
{Savageau}, M.~A. (1976).
\newblock {\em {Biochemical systems analysis. A study of function and design in
  molecular biology}\/}, vol. 6739 of {\em {Advanced Book Program}\/}.
\newblock {London etc.}: Addison-Wesley Publishing Company.

\bibitem[{{Schittkowski}(2002)}]{schittkowski02}
{Schittkowski}, K. (2002).
\newblock {\em {Numerical data fitting in dynamical systems. A practical
  introduction with applications and software}\/}.
\newblock Dordrecht: Kluwer Academic Publishers.

\bibitem[{Tufte(2001)}]{tufte01}
Tufte, E.~R. (2001).
\newblock {\em The visual display of quantitative information\/}.
\newblock Cheshire, CT: Graphics Press, 2nd ed.

\bibitem[{Varah(1982)}]{varah1982spline}
Varah, J. (1982).
\newblock A spline least squares method for numerical parameter estimation in
  differential equations.
\newblock {\em SIAM Journal on Scientific and Statistical Computing\/}, {\em
  3\/}(1), 28--46.

\bibitem[{{Venables} \& {Ripley}(2002)}]{ripley2002}
{Venables}, W.~N., \& {Ripley}, B.~D. (2002).
\newblock {\em {Modern applied statistics with S}\/}.
\newblock {Statistics and Computing}. New York, NY: Springer, 4th ed.

\bibitem[{Vilela et~al.(2007)Vilela, Borges, Vinga, Vasconcelos, Santos, Voit,
  \& Almeida}]{vilela2007}
Vilela, M., Borges, C. C.~H., Vinga, S., Vasconcelos, A. T.~R., Santos, H.,
  Voit, E.~O., \& Almeida, J.~S. (2007).
\newblock Automated smoother for the numerical decoupling of dynamics models.
\newblock {\em BMC Bioinformatics\/}, {\em 8\/}(1), 305.

\bibitem[{{Vissing Mikkelsen} \& {Hansen}(2017)}]{mikkelsen2017learning}
{Vissing Mikkelsen}, F., \& {Hansen}, N.~R. (2017).
\newblock {Learning large scale ordinary differential equation systems}.
\newblock {\em arXiv e-prints\/}.
\newline\urlprefix\url{https://arxiv.org/abs/1710.09308}

\bibitem[{Voit(2000)}]{voit2000computational}
Voit, E.~O. (2000).
\newblock {\em Computational analysis of biochemical systems: A practical guide
  for biochemists and molecular biologists\/}.
\newblock Cambridge University Press.

\bibitem[{Voit(2013)}]{voit2013biochemical}
Voit, E.~O. (2013).
\newblock Biochemical systems theory: a review.
\newblock {\em ISRN Biomathematics\/}, {\em 2013\/}, Article ID 897658.

\bibitem[{{Volterra}(1926)}]{volterra1926}
{Volterra}, V. (1926).
\newblock {Fluctuations in the abundance of a species considered
  mathematically}.
\newblock {\em {Nature}\/}, {\em 118\/}, 558--560.

\bibitem[{Vuja{\v{c}}i{\'c} et~al.(2015)Vuja{\v{c}}i{\'c}, Dattner,
  Gonz{\'a}lez, \& Wit}]{vujavcic2015time}
Vuja{\v{c}}i{\'c}, I., Dattner, I., Gonz{\'a}lez, J., \& Wit, E. (2015).
\newblock Time-course window estimator for ordinary differential equations
  linear in the parameters.
\newblock {\em Stat. Comput.\/}, {\em 25\/}(6), 1057--1070.

\bibitem[{{Wasserman}(2006)}]{wasserman06nonparametric}
{Wasserman}, L. (2006).
\newblock {\em {All of nonparametric statistics}\/}.
\newblock {Springer Texts in Statistics}. New York, NY: Springer.

\bibitem[{{Wickham}(2009)}]{wickham09}
{Wickham}, H. (2009).
\newblock {\em {ggplot2. Elegant graphics for data analysis}\/}.
\newblock {Use R!} New York, NY: Springer.

\bibitem[{Wu et~al.(2019)Wu, Qiu, Yuan, \& Wu}]{wu2018parameter}
Wu, L., Qiu, X., Yuan, Y.-x., \& Wu, H. (2019).
\newblock Parameter estimation and variable selection for big systems of linear
  ordinary differential equations: A matrix-based approach.
\newblock {\em J. Amer. Statist. Assoc.\/}, {\em 114\/}(526), 657--667.

\bibitem[{{Yaari} \& {Dattner}(2018)}]{yaari2018textbf}
{Yaari}, R., \& {Dattner}, I. (2018).
\newblock {simode: R Package for statistical inference of ordinary differential
  equations using separable integral-matching}.
\newblock {\em arXiv e-prints\/}.
\newline\urlprefix\url{https://arxiv.org/abs/1807.04202}

\bibitem[{Yaari \& Dattner(2019)}]{simode}
Yaari, R., \& Dattner, I. (2019).
\newblock {\em simode: Statistical inference for systems of ordinary
  differential equations using separable integral-matching\/}.
\newblock R package version 1.1.4.
\newline\urlprefix\url{https://CRAN.R-project.org/package=simode}

\bibitem[{Yaari et~al.(2018)Yaari, Dattner, \& Huppert}]{yaarietal18}
Yaari, R., Dattner, I., \& Huppert, A. (2018).
\newblock A two-stage approach for estimating the parameters of an age-group
  epidemic model from incidence data.
\newblock {\em Stat. Methods Med. Res.\/}, {\em 27\/}(7), 1999--2014.

\end{thebibliography}

\end{document}